\documentclass[12pt]{article}

\usepackage{scicite}

\usepackage{times}
\usepackage{graphicx}
\usepackage{amssymb}

\newenvironment{sciabstract}{%
\begin{quote} \bf}
{\end{quote}}

\newcounter{lastnote}

\title{Gamma-ray flares from the Crab Nebula} 
\date{}

\begin{document} 



\maketitle 

\noindent
A.~A.~Abdo$^{1}$, 
M.~Ackermann$^{2}$, 
M.~Ajello$^{2}$, 
A.~Allafort$^{2}$, 
L.~Baldini$^{3}$, 
J.~Ballet$^{4}$, 
G.~Barbiellini$^{5,6}$, 
D.~Bastieri$^{7,8}$, 
K.~Bechtol$^{2}$, 
R.~Bellazzini$^{3}$
% ...
\medskip
\begin{enumerate}
\item[1.] National Research Council Research Associate, National Academy of Sciences, Washington, DC 20001, resident at Naval Research Laboratory, Washington, DC 20375, USA
\item[2.] W. W. Hansen Experimental Physics Laboratory, Kavli Institute for Particle Astrophysics and Cosmology, Department of Physics and SLAC National Accelerator Laboratory, Stanford University, Stanford, CA 94305, USA
\item[3.] Istituto Nazionale di Fisica Nucleare, Sezione di Pisa, I-56127 Pisa, Italy
\item[4.] Laboratoire AIM, CEA-IRFU/CNRS/Universit\'e Paris Diderot, Service d'Astrophysique, CEA Saclay, 91191 Gif sur Yvette, France
\item[5.] Istituto Nazionale di Fisica Nucleare, Sezione di Trieste, I-34127 Trieste, Italy
\item[6.] Dipartimento di Fisica, Universit\`a di Trieste, I-34127 Trieste, Italy
\item[7.] Istituto Nazionale di Fisica Nucleare, Sezione di Padova, I-35131 Padova, Italy
\item[8.] Dipartimento di Fisica ``G. Galilei", Universit\`a di Padova, I-35131 Padova, Italy
\item[9.] Istituto Nazionale di Fisica Nucleare, Sezione di Perugia, I-06123 Perugia, Italy
\item[10.] Dipartimento di Fisica, Universit\`a degli Studi di Perugia, I-06123 Perugia, Italy
\item[11.] Centre d'\'Etude Spatiale des Rayonnements, CNRS/UPS, BP 44346, F-30128 Toulouse Cedex 4, France
\item[12.] Department of Physics, Center for Cosmology and Astro-Particle Physics, The Ohio State University, Columbus, OH 43210, USA
\item[13.] Dipartimento di Fisica ``M. Merlin" dell'Universit\`a e del Politecnico di Bari, I-70126 Bari, Italy
\item[14.] Istituto Nazionale di Fisica Nucleare, Sezione di Bari, 70126 Bari, Italy
\item[15.] Laboratoire Leprince-Ringuet, \'Ecole polytechnique, CNRS/IN2P3, Palaiseau, France
\item[16.] Institut de Ciencies de l'Espai (IEEC-CSIC), Campus UAB, 08193 Barcelona, Spain
\item[17.] NASA Goddard Space Flight Center, Greenbelt, MD 20771, USA
\item[18.] University College Dublin, Belfield, Dublin 4, Ireland
\item[19.] INAF-Istituto di Astrofisica Spaziale e Fisica Cosmica, I-20133 Milano, Italy
\item[20.] Center for Research and Exploration in Space Science and Technology (CRESST) and NASA Goddard Space Flight Center, Greenbelt, MD 20771, USA
\item[21.] Department of Physics and Center for Space Sciences and Technology, University of Maryland Baltimore County, Baltimore, MD 21250, USA
\item[22.] College of Science, George Mason University, Fairfax, VA 22030, resident at Naval Research Laboratory, Washington, DC 20375, USA
\item[23.] Laboratoire de Physique Th\'eorique et Astroparticules, Universit\'e Montpellier 2, CNRS/IN2P3, Montpellier, France
\item[24.] Agenzia Spaziale Italiana (ASI) Science Data Center, I-00044 Frascati (Roma), Italy
\item[25.] IASF Palermo, 90146 Palermo, Italy
\item[26.] INAF-Istituto di Astrofisica Spaziale e Fisica Cosmica, I-00133 Roma, Italy
\item[27.] Space Science Division, Naval Research Laboratory, Washington, DC 20375, USA
\item[28.] Dipartimento di Fisica, Universit\`a di Udine and Istituto Nazionale di Fisica Nucleare, Sezione di Trieste, Gruppo Collegato di Udine, I-33100 Udine, Italy
\item[29.] Istituto Universitario di Studi Superiori (IUSS), I-27100 Pavia, Italy
\item[30.] Universit\'e Bordeaux 1, CNRS/IN2p3, Centre d'\'Etudes Nucl\'eaires de Bordeaux Gradignan, 33175 Gradignan, France
\item[31.] Osservatorio Astronomico di Trieste, Istituto Nazionale di Astrofisica, I-34143 Trieste, Italy
\item[32.] Department of Physical Sciences, Hiroshima University, Higashi-Hiroshima, Hiroshima 739-8526, Japan
\item[33.] INAF Istituto di Radioastronomia, 40129 Bologna, Italy
\item[34.] Center for Space Plasma and Aeronomic Research (CSPAR), University of Alabama in Huntsville, Huntsville, AL 35899, USA
\item[35.] Science Institute, University of Iceland, IS-107 Reykjavik, Iceland
\item[36.] Department of Physics and Department of Astronomy, University of Maryland, College Park, MD 20742, USA
\item[37.] Research Institute for Science and Engineering, Waseda University, 3-4-1, Okubo, Shinjuku, Tokyo, 169-8555 Japan
\item[38.] Department of Physics, University of Washington, Seattle, WA 98195-1560, USA
\item[39.] Istituto Nazionale di Fisica Nucleare, Sezione di Roma ``Tor Vergata", I-00133 Roma, Italy
\item[40.] Department of Physics and Astronomy, University of Denver, Denver, CO 80208, USA
\item[41.] Hiroshima Astrophysical Science Center, Hiroshima University, Higashi-Hiroshima, Hiroshima 739-8526, Japan
\item[42.] Institute of Space and Astronautical Science, JAXA, 3-1-1 Yoshinodai, Chuo-ku, Sagamihara, Kanagawa 252-5210, Japan
\item[43.] Institut f\"ur Astro- und Teilchenphysik and Institut f\"ur Theoretische Physik, Leopold-Franzens-Universit\"at Innsbruck, A-6020 Innsbruck, Austria
\item[44.] Santa Cruz Institute for Particle Physics, Department of Physics and Department of Astronomy and Astrophysics, University of California at Santa Cruz, Santa Cruz, CA 95064, USA
\item[45.] Space Sciences Division, NASA Ames Research Center, Moffett Field, CA 94035-1000, USA
\item[46.] NYCB Real-Time Computing Inc., Lattingtown, NY 11560-1025, USA
\item[47.] Department of Chemistry and Physics, Purdue University Calumet, Hammond, IN 46323-2094, USA
\item[48.] Partially supported by the International Doctorate on Astroparticle Physics (IDAPP) program
\item[49.] Instituci\'o Catalana de Recerca i Estudis Avan\c{c}ats (ICREA), Barcelona, Spain
\item[50.] Consorzio Interuniversitario per la Fisica Spaziale (CIFS), I-10133 Torino, Italy
\item[51.] INTEGRAL Science Data Centre, CH-1290 Versoix, Switzerland
\item[52.] NASA Postdoctoral Program Fellow, USA
\item[53.] Dipartimento di Fisica, Universit\`a di Roma ``Tor Vergata", I-00133 Roma, Italy
\item[54.] Department of Physics, Stockholm University, AlbaNova, SE-106 91 Stockholm, Sweden
\item[55.] The Oskar Klein Centre for Cosmoparticle Physics, AlbaNova, SE-106 91 Stockholm, Sweden
\item[$\dagger$] buehler@stanford.edu, * rdb3@stanford.edu , $\ddagger$funk@slac.stanford.edu
\end{enumerate}

\clearpage


\begin{sciabstract}
  A young and energetic pulsar powers the well-known Crab Nebula.
  Here we describe two separate gamma-ray (photon energy $>$100 MeV)
  flares from this source detected by the Large Area Telescope on
  board the Fermi Gamma-ray Space Telescope. The first flare occurred
  in February 2009 and lasted approximately 16 days. The second flare
  was detected in September 2010 and lasted approximately 4
  days. During these outbursts the gamma-ray flux from the nebula
  increased by factors of four and six, respectively. The brevity of
  the flares implies that the gamma rays were emitted via synchrotron
  radiation from PeV ($10^{15}$\,eV) electrons in a region smaller
  than 1.4 $\times$ 10$^{-2}$ pc.  These are the highest energy
  particles that can be associated with a discrete astronomical
  source, and they pose challenges to particle acceleration theory.
\end{sciabstract}

The Crab Nebula is the remnant of an historical supernova (SN),
recorded in 1054 C.E., located at a distance of 2 kpc
\cite{trimble_distance_1973}. The SN explosion left behind a pulsar,
which continuously emits a wind of magnetized plasma of
electron/positron pairs (henceforth referred to as electrons). This
pulsar wind is expected to terminate in a standing shock where the
particles may undergo shock acceleration
\cite{rees_origin_1974,kennel_magnetohydrodynamic_1984}. As the
electrons diffuse into the downstream medium they release energy
through interactions with the surrounding magnetic and photon
fields. This emission is observed across all wavebands from radio up
to TeV gamma-ray energies and is referred to as a pulsar wind nebula
(PWN). The efficiency of this process is remarkable. As much as 30\%
of the total energy released by the Crab pulsar is emitted by the PWN
[\cite{hester_crab_2008} and references therein].
The Crab PWN has an approximately ellipsoidal shape on the sky with a
size that decreases with increasing photon energy. At radio
frequencies it extends out to 5$^{\prime}$ (3 pc) from the central
pulsar. At X-ray wavelengths a bright torus surrounds the pulsar; its
radius is 40$^{\prime\prime}$ (0.4 pc) and jets emerge perpendicular
to it in both directions.

Within the region encapsulated by the torus there are several
small-scale structures. The inner nebula, which we define as the
central 15$^{\prime\prime}$ around the pulsar, has several small-scale
regions of variable X-ray and optical brightness. The most prominent
is an X-ray-bright inner ring with a radius of 10$^{\prime\prime}$
(0.1 pc); this ring is thought to represent the termination shock of
the PWN \cite{weisskopf_discovery_2000}. Several knots with diameters
of $\sim$1$^{\prime\prime}$ (0.01 pc) are detected close to the inner
ring and the base of the jets, and bright arcs of comparable width are
observed moving outwards from the inner ring into the
torus\cite{scargle_activity_1969,hester_hubble_2002}.

The broad-band spectral energy distribution (SED) of the Crab Nebula
is composed of two broad non-thermal components. A low-energy
component dominates the overall output and extends from radio to
gamma-ray frequencies.  This emission is thought to be from
synchrotron radiation. This notion is confirmed in radio to X-ray
frequencies with polarization measurements \cite{cocke_optical_1970,
  novick_detection_1972,dean_polarized_2008}. The emission of this
synchrotron component peaks between optical and X-ray frequencies,
where the emission is primarily from the
torus\cite{weisskopf_discovery_2000}.  The emission site of
higher-energy photons (beyond 100 keV) cannot be resolved due to the
limited angular resolution of telescopes observing at these
frequencies. The high-energy component dominates the emission above
$\sim$ 400 MeV and is thought to be emitted via inverse Compton (IC)
scattering, predominantly of the synchrotron
photons\cite{gould_high_1965,atoyan_mechanisms_1996}.

The large-scale integrated emission from the Crab Nebula is expected
to be steady within a few percent and is thus often used to
cross-calibrate X-ray and gamma-ray telescopes and to check their
stability over time \cite{weisskopf_calibrations_2010 ,
  meyer_crab_2010}. Recently, variability in the x-ray flux from the
nebula by $\sim$3.5\% yr$^{-1}$ has been detected, setting limits on
the accuracy of this practice \cite{wilson-hodge_whenstandard_2010}.
Yearly variations in the emission in the high-energy tail (1-150 MeV)
of the synchrotron component has also been
reported\cite{much_comptel_1995,de_jager_gamma-ray_1996}. No
significant variations have been detected for the high-energy
component of the nebula
\cite{aharonian_crab_2004,aharonian_observations_2006,albert_vhe_2008}.

The Large Area Telescope (LAT) on board the Fermi Gamma-Ray Space
Telescope (Fermi) has continuously monitored the Crab Nebula as a part
of its all-sky survey since August 2008. The LAT detects gamma rays
from 20 MeV to $>$300 GeV, and this spans the transition region
between the low and the high-energy components of the nebular
spectrum. The average SED measured during the first 25 months of
observations (Fig. 1) is well characterized by the sum of two spectral
components, each with a power-law dependence on energy
\cite{abdo_fermi_2010}. The integrated flux of the low-energy
component is (6.2 $\pm$ 0.3) $\times$ 10$^{-7}$ cm$^{-2}$ above 100
MeV with an photon index of 3.69 $\pm$ 0.11 (only statistical errors
are given; see online supplements for a discussion of systematic
errors). The high-energy component has an integral flux of (1.3 $\pm$
0.1) $\times$ 10$^{-7}$ cm$^{-2}$ s$^{-1}$ above 100 MeV with a photon
index of 1.67 $\pm$ 0.04. Due to its hard energy spectrum the
high-energy component dominates the emission above 426 $\pm$ 35 MeV.

In order to search for flux variability of both spectral components in
the LAT band, we grouped the flux measurements into monthly time
bins. The high-energy component was found to be stable. The low-energy
component was found to vary on these time scales (Fig. 2); the
probability that the measured flux variations are statistical
measurement fluctuations in a constant source is less than
10$^{-5}$. No significant spectral variations were detected for either
component on monthly time scales.  Flux variability was also searched
for on sub-monthly time scales, for which the low-energy component of
the nebula is significantly detected by the LAT only in high-flux
states.  The flux of the low-energy component was significantly
enhanced compared to the average values in February 2009 and September
2010 (Fig. 2).  No variations were found for the high-energy
component. The September flare was first announced by the AGILE
gamma-ray mission \cite{atel_agile}, which additionally
reports a flare in October 2007, before the start of Fermi
observations\cite{agile_paper}.  The Fermi-LAT detected  flare in
February 2009 was not detected by AGILE as the instrument was pointing
at a different part of the sky. 

The February flare had a duration of $\sim$16 days. The average
integral flux above 100 MeV of the low-energy component between MJD
54857.73 and 54873.73 was (23.2 $\pm$ 2.9) $\times$ 10$^{-7}$ cm$^{-2}$
s$^{-1}$, corresponding to an increase by a factor 3.8 $\pm$ 0.5
compared with the average value; the increase is significant at
$>$8-$\sigma$ level. The September flare lasted for only $\sim 4$ days. The
integral flux above 100 MeV between MJD 55457.73 and 55461.73 was (33.8
$\pm$ 4.6) $\times$ 10$^{-7}$ cm$^{-2}$ s$^{-1}$, corresponding to an 
increase by a factor 5.5 $\pm$ 0.8 with respect to the average and a
significance of $>$10 $\sigma$. 
 
The February flare has a soft spectrum with a photon index of 4.3
$\pm$ 0.3 (Fig. 1). The spectral slope is compatible with the average
25-month value within two standard deviations. The energy spectrum for
the second flare was significantly harder, with a photon index of 2.7
$\pm$ 0.2, and was still detected above 1 GeV at a
3-$\sigma$-level. The average power released in each of the gamma-ray
flares was approximately 4 $\times 10^{36}$ erg s$^{-1}$, for the case
of isotropic emission. No significant variations in the emission of
the pulsar were detected on monthly and four-day time scales through
the period of observations. Examination of the timing residuals of the
pulsed emission indicated no significant variations during either
flares nor any significant glitch activity during the first 25 months
of LAT observations.

No variations in the synchrotron component between infrared and X-ray
frequencies were seen about the average nebular flux level during the
second flare \cite{atels_irx}. We analyzed data collected by the BAT
instrument on board the Swift satellite \cite{ajello_swift_2008},
which continuously monitors the sky at photon energies of 15$-$150
keV. The mean flux measured during the first flare was (2.0 $\pm$ 0.1)
$\times$ 10$^{-8}$ erg cm$^{-2}$ s$^{-1}$, the flux during the second
flare was (2.0 $\pm$ 0.1) $\times$ 10$^{-8}$ erg cm$^{-2}$ s$^{-1}$.
Both observations are therefore within 5\% of the average flux of
(2.09$\pm$ 0.10) $\times$ 10$^{-8}$ erg cm$^{-2}$ s$^{-1}$ measured by
BAT in this energy range \cite{batdigest}, and show no correlation to
the gamma-ray flares. The angular resolution of the BAT only allows
for the measurement of the spatially integrated
spectrum. Sub-arc-second resolution images were taken in X-rays by the
Chandra observatory and optical by the Hubble Space Telescope a few
days after the second flare. Although both images show no unusual
activity compared to previous observations, both show a brightening
3$^{\prime\prime}$ east of the pulsar \cite{atels_chandrahubble}.  In
the Chandra image this brightening is associated with a knot of
$\sim$1$^{\prime\prime}$ diameter that might be associated with the
inner ring or the base of the jet.  Such a brightening might be
interpreted as an afterglow at lower frequencies of the gamma-ray
flare, but no conclusions can be drawn based on one event.

The brief flare time scales and the requirement that the emission
volume be causally connected imply that the flaring region must have
been compact.  If $L$ is the diameter of the flaring region along the
line-of-sight and $t$ is the flare duration, then $L < Dct$, where the
Doppler factor $D$ accounts for relativistic boosting effects.  The
Doppler factor is expected to be moderate within the Crab Nebula, as
the typical velocities observed are smaller than 0.9 $c$
\cite{hester_hubble_2002}. Even if the emission region was moving
directly toward us, this yields $D<$4.4. For a flare duration of 4
days this results in $L$ $<$ 1.4 $\times$ 10$^{-2}$ pc, which
corresponds to $<$1.5 $^{\prime\prime}$ projected on the
sky. Structures this small are found only in the inner part of the
nebula, close to the termination shock, the base of the jet or the
pulsar, suggesting that the gamma-ray emission detected in the flare
originated from these regions. This is in agreement with expectations
of relativistic magnetohydrodynamic simulations, in which the
gamma-ray emission of the synchrotron component originates close to
the termination shock
\cite{volpi_non-thermal_2008,komissarov_origin_2010}.


The extrapolation of the the LAT spectrum of low-energy component to
lower frequencies suggest that it represents synchrotron emission
(Fig. 1). The brevity of the gamma-ray flares strengthens this
scenario: If the flare were instead produced by IC radiation or
Bremsstrahlung, the cooling time of the emitting electrons would
greatly exceed the flare duration. The cooling via Bremsstrahlung in
particle densities $<$10 cm$^{-3}$ \cite{atoyan_fluxes_1996} happens
over $\sim$10$^6$ years.  Similarly, electrons cooling via IC emission
of 100 MeV gamma rays on the photons of the synchrotron component of
Crab Nebula have cooling times $\gtrsim$ 10$^7$ years.  The average
magnetic field inside the Crab Nebula is estimated to be $\sim$200
$\mu$G, as deduced from modeling of the broad-band SED
\cite{atoyan_mechanisms_1996, abdo_fermi_2010}, and might be enhanced
locally by up to an order of magnitude in the inner
nebula\cite{hester_wfpc2_1995}. These fields imply synchrotron cooling
times $\lesssim$15 days, comparable to the flare duration, leaving
synchrotron radiation as the only plausible process responsible for
the gamma-ray emission during the flares.

The detection of synchrotron photons up to energies of $>$1 GeV
confirms that electrons are accelerated to energies of $\gtrsim$1 PeV
in the Crab Nebula \cite{de_jager_expected_1992}.  These are the
highest energy particles that can be associated directly with any
astronomical source, and they pose special challenges to particle
acceleration theory. Because synchrotron losses are so efficient,
there must be a strong electric field $E$ to compensate radiation
reaction, given by:
\begin{equation}
E/B\approx r_{{\rm L}}/\ell_{{\rm cool}}\gtrsim(1.3 \alpha{\mathcal E}_{\gamma\rm{pk}}/m_ec^2)\approx ({\mathcal E}_{\gamma{\rm pk}}/50{\rm MeV})
\end{equation} 
where $r_{{\rm L}}$ is the Larmor radius, $\ell_{{\rm cool}}$ is the
radiative cooling length, $\alpha$ is the fine structure constant and
${\mathcal E}_{\gamma\rm{pk}}$ is the peak synchrotron frequency at
which the most energetic electrons are emitting
\cite{Lyutikov_2010,de_jager_gamma-ray_1996}. Due to the detection of
gamma-ray emission beyond 1 GeV ${\mathcal E}_{\gamma\rm{pk}}$ can be
conservatively estimated to be greater than 200 MeV. The electric
field is unlikely to exceed the magnetic field; if it did, there would
be a local reference frame with a pure electric field in which vacuum
breakdown would occur quickly. We conclude that the electric field, as
measured in the Crab frame, is close in magnitude to the magnetic
field in the region where the highest energy synchrotron photons were
emitted. This subsumes the possibility of bulk relativistic
motion. Furthermore, the resistive force due to radiation reaction is
competitive with the Lorentz force and the cooling length is
comparable with the Larmor radius. This poses severe difficulties to
the widely-discussed acceleration mechanism of diffusive shock
acceleration\cite{gallant_particle_2002,sironi_particle_2009}. The
proposed acceleration due to absorption of ion cyclotron waves does
not suffer from these constraints \cite{amato_heating_2006}. However,
it appears to operate on time scales which are too long to accommodate
the fast variability seen during the flares. Alternatively, the
acceleration could be related directly to the electric field from the
pulsar.

The Crab Nebula is powered by the central neutron star which acts as a
DC unipolar inductor and a source of an AC striped
wind\cite{rees_origin_1974,kennel_magnetohydrodynamic_1984}. What
happens to the DC and AC current flows is controversial. It is widely
supposed that $\sim90$\% of the DC current returns in an outflowing
wind that becomes particle-dominated and encounters a (mostly
invisible) termination shock at a radius
$\sim0.1$~pc\cite{Buccianti_2010}, but the wind could also remain
electromagnetically dominated \cite{Blandford_2002,Lyutikov_2010}. For
the measured spin-down rate, a moment of inertia of
$\sim1\times10^{45}$~g cm$^{2}$ and a force-free model of the
magnetosphere, the total induced potential difference is $\sim50$~PV,
high enough to accelerate particles to the required energies. The
current associated with this potential is $\sim300$~TA yielding a DC
power per hemisphere of $\sim1.5\times10^{38}$~erg s$^{-1}$, a factor
$\sim$40 larger than the power released in the flares. Another
interesting possibility is that particle acceleration takes place in
the AC striped wind of the pulsar due to magnetic reconnection,
although it is not clear if this process can accelerate particles to
PeV energies on the required time scales
\cite{lyubarsky_particle_2008,bednarek_variability_2010}.

The observations reported here have raised compelling questions on our
understanding of particle acceleration and motivate more detailed
calculations; together with the ongoing gamma-ray observations of the
LAT and observational campaigns at X-ray and optical wavelengths they
might soon pinpoint the gamma-ray emission site in the Crab Nebula.

\clearpage

\begin{figure}[!t]
\centering
\includegraphics[width=14cm]{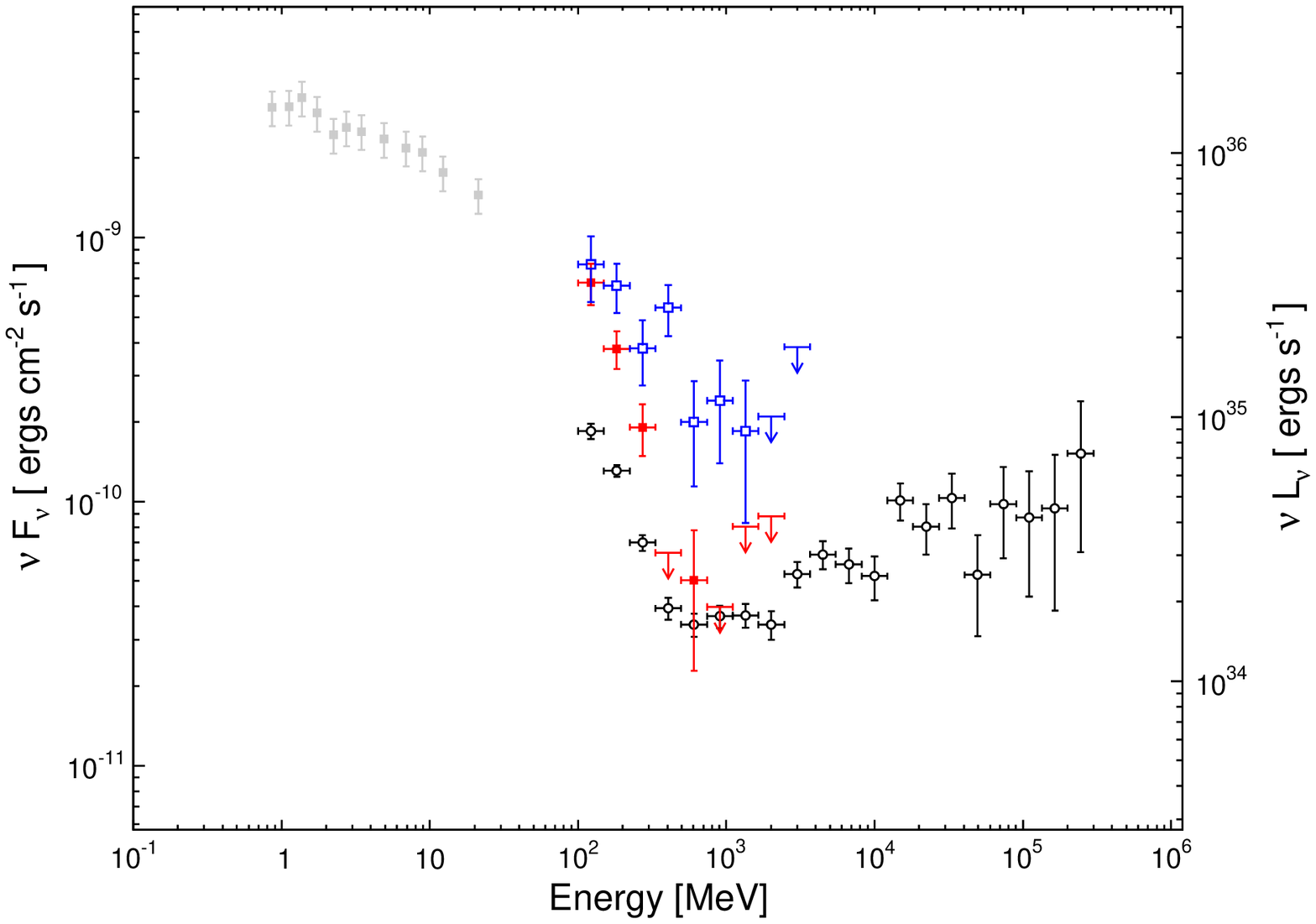}
\caption{Spectral energy distribution of the Crab Nebula. Black open
  circles show the average spectrum measured by the LAT in the first
  25 months of observations. Red squares show the energy spectrum during
  the flare of February 2009 (MJD 54857.73-54873.73) and blue open
  squares the spectrum in September 2010 (MJD 55457.73-55461.73). Gray
  squares show historical long-term average spectral data from the
  COMPTEL telescope with 15\% systematic errors
  \cite{kuiper_crab_2001}.  Arrows indicate 95\% confidence flux
  limits.}
\label{fig::sed}
\end{figure}

\clearpage

\begin{figure}[!t]
\centering
\includegraphics[width=15cm]{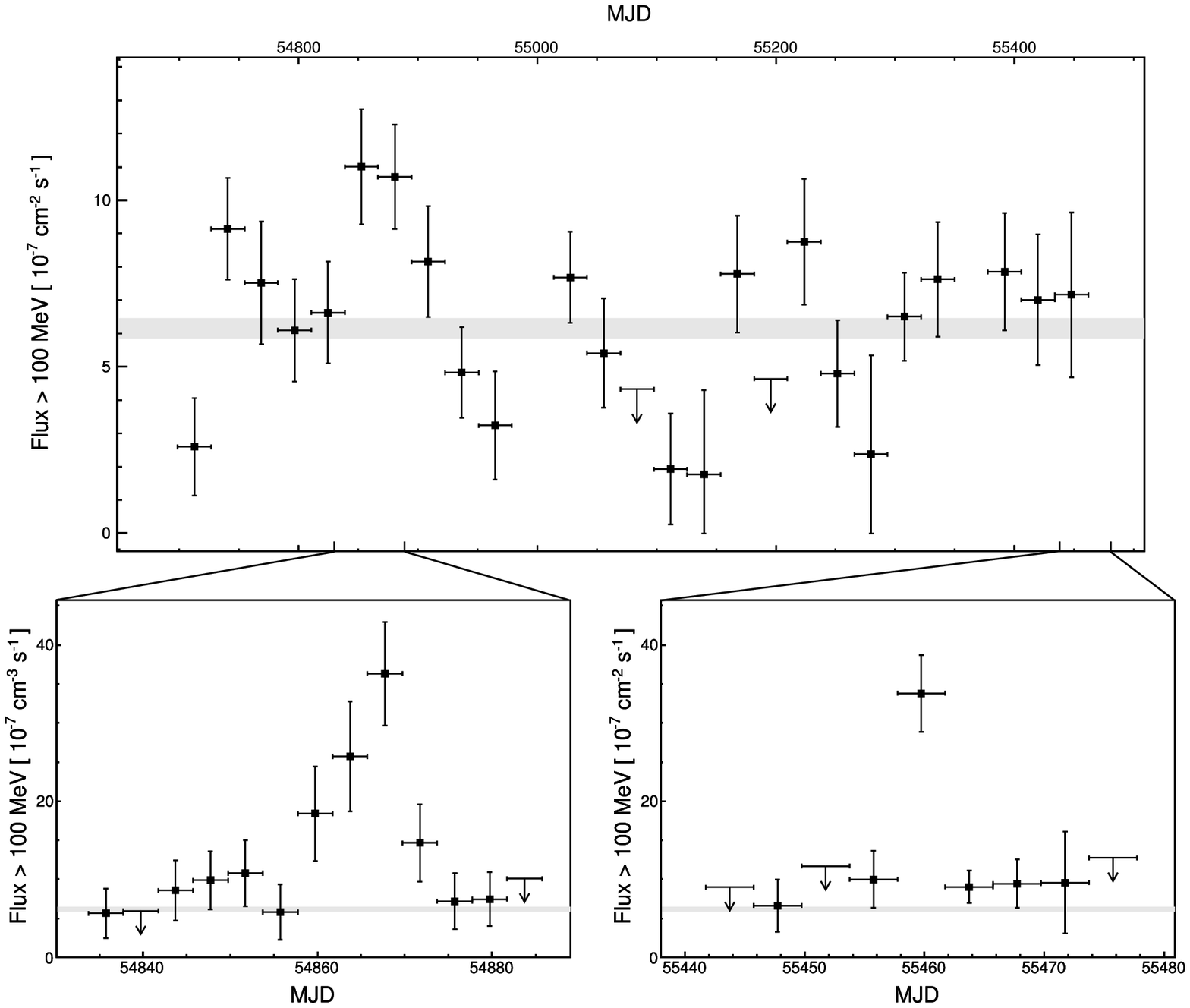}
\caption{Gamma-ray flux above 100 MeV as a function of time of the
  synchrotron component of the Crab Nebula. The upper panel shows the
  flux in four-week intervals for the first 25 month of
  observations. Data for times when the sun was within 15$^{\circ}$ of the Crab
  Nebula have been omitted. The gray band indicates the average flux
  measured over the entire period. The lower panel shows the flux as a
  function of time in four-day time bins during the flaring periods in
  February 2009 and September 2010. Arrows indicate 95\% confidence
  flux limits.} 
\label{fig::lc}
\end{figure}

\clearpage

\bibliography{crab}

\bibliographystyle{Science}

\clearpage

\section*{Supplements}

\subsection*{Fermi-LAT Analysis}

The LAT analysis presented in this publication was performed using the
Science Tools v9r17p0 with the P6\_V3\_DIFFUSE instrument response
functions \cite{fssc}. Fluxes were estimated by maximizing the
likelihood of a given source model using the unbinned gtlike tool. The
model was composed of the Galactic and isotropic diffuse models (v02
\cite{latcatalog}), as well as all sources within 15$^{\circ}$ of the
Crab pulsar in the first Fermi LAT catalog (1FGL
\cite{latcatalog}). Only photons within this same radius and with
energies greater than 100 MeV were considered. All data taken under
optimal conditions between MJD 54684.73 and 55477.73 were considered,
except for periods when the sun was within 15$^\circ$ of the Crab
Nebula. To avoid background contamination from the gamma-ray emission
of the earth limb only time periods where the Crab Nebula was at a
zenith angle $<$90$^{\circ}$ were considered. In order to avoid the
strong foreground emission of the pulsar, only rotational phases
between 0.52 and 0.87 were considered for the nebula analysis
\cite{abdo_fermi_2010}. The pulsar timing model used will be made
available through the Fermi Science Support Center \cite{fssc}.

Light curves were generated by performing the likelihood fit in each
time bin independently. The free parameters in the 28-day likelihood
fits were the fluxes and spectral indices of the synchrotron and IC
components of the nebula and the amplitude of the Galactic and
isotropic diffuse emission.  All other parameters were held fixed to
their 1FGL values.  In the 4-day light curves, the only free
parameters of the likelihood fits were the flux and spectral index of
the synchrotron component. The IC flux and spectral index and the
Galactic and isotropic diffuse normalizations were held fixed to their
25-month best-fit values.  Spectral data points on the SED were
measured by performing the likelihood fit in each energy bin
independently, varying only the normalization of the best-fit
model. For both light curves and energy spectra, 95\% confidence upper
limits on the flux were derived if the significance of the source
detection was below 2$\sigma$.

\subsubsection*{Significance and position of the flares}

The flaring periods were identified by fitting a flaring component
together with the average 25-month spectral model in weekly time bins
between MJD 54684.73 and 55461.73. The average model included the two
components of the Crab Nebula.  The flaring component was modeled by a
power-law energy spectrum and its normalization and index were the
only parameters varied in the fits. The significance of the flaring
component was $<$3.5 $\sigma$ for all weeks outside of the flare
periods, while it was $>$6.7 $\sigma$ during the flares. Considering
trials for the 102 analyzed weeks this corresponds to significances
$<$2 and $>$6 $\sigma$ respectively.  To measure the properties of the
emission during the flare in more detail two flaring periods were
defined from the 4 day light curves shown in Figure 2. For the first
flare the flaring period was defined as MJD 54857.73-54873.73 (the
four high flux bins around the flare). Fitting the additional source
with the average best-fit model yields a significance of $>$8 $\sigma$
for the flaring component. For the second flare the flaring period was
defined as MJD 55457.73-55461.73. The significance of the flaring
component during this time period is $>$10~$\sigma$.

For both flares the position of the flaring component was derived
using the gtfindsrc program of the Science Tools. Both positions are
compatible with the Crab Nebula. The best-fit position of the first
flare is R.A.: 83.73$^{\circ}$, Dec: 21.81$^{\circ}$ with a 68\%
containment radius of 0.22$^{\circ}$ (J2000), one standard deviation
away from the Crab pulsar position of R.A.: 83.63$^{\circ}$ Dec:
22.01$^{\circ}$.  The best-fit position of the second flare is R.A.:
83.68$^{\circ}$, Dec: 22.03$^{\circ}$ with a 68\% containment radius
of 0.044$^{\circ}$, 1.1 $\sigma$ away from the pulsar.

\subsubsection*{The possibility of a background blazar}

The monthly flux variability and flaring periods in principle might be
due to the emission of a background source that cannot be resolved
within the angular resolution of the LAT. However, this scenario is
very unlikely for several reasons. First, the flux increase during
both flares above 100 MeV was $\gtrsim$ 2 $\times$ 10$^{-6}$
cm$^{-2}$; there are fewer than 100 LAT sources which show variability
of this magnitude. The probability of chance coincidence for one of
them to be located behind the Crab Nebula within the 2 $\sigma$
localization error measured for the second flare, is $<$6 $\times$
10$^{-5}$.  Second, the spectral index during the first flare is the
softest of any source yet detected by the LAT, with the exception of
the synchrotron component of the Crab Nebula itself
\cite{latcatalog,abdo_fermi_2010}, suggesting that the emission
originate from the same source (this argument also relates the second
flare to the Crab Nebula, as the probability that both flares are
produced by two different sources is negligible, due to the even lower
chance probability of positional coincidence in this case).  Finally,
the only source class known to produce variability on the flare time
scales of the observed magnitude in gamma-rays are blazars. No known
blazar is located near the Crab Nebula within the angular resolution
of Fermi and X-ray observations taken two days after the second flare
revealed no new source which might be associated with a yet-unknown
blazar \cite{atel_blazar}.

\subsubsection*{Systematic Uncertainties}

Stability of flux measurements with the LAT data over time was tested
using sources which are expected to be constant in flux. These sources
are the Crab pulsar, the nearby Geminga pulsar, the Vela pulsar and
the Galactic diffuse emission. We found the light curve of each of
these reference sources was consistent with a constant on monthly and
4-day timescales if we added a 5\% systematic error in quadrature with
the statistical errors.  We have therefore added a 5\% systematic
error in quadrature to all light curves in this publication and have
taken this systematic error to be time-variable and normally
distributed.  We adopt 30\% as the systematic uncertainty for overall
normalization of integrated fluxes $>$100 MeV reported here.  The
errors given on LAT spectral points are statistical only. The
uncertainties in the energy dependence of the acceptance and energy
disperson of the LAT leads to a systematic error of 0.1 on the photon
index \cite{abdo_fermi_2010}.

\end{document}